\documentclass[preprint]{revtex4}
\usepackage{amsmath}
\usepackage{float}
\usepackage{graphics,epsfig}
\usepackage{graphicx}
\usepackage{dcolumn}
\usepackage{bm}

\begin{document}
\baselineskip=0.8 cm
\title{{\bf Gauss-Bonnet Term on Vacuum Decay}}
\author{Rong-Gen Cai}
\email{cairg@itp.ac.cn} \affiliation{Institute of Theoretical
Physics, Chinese Academy of Sciences, P.O.Box 2735, Beijing 100190,
China}

\author{Bin Hu}
\email{hubin@itp.ac.cn} \affiliation{Institute of Theoretical
Physics, Chinese Academy of Sciences, P.O.Box 2735, Beijing 100190,
China\\
Graduate University of Chinese Academy of Sciences, YuQuan Road 19A,
Beijing 100049, China}

\author{Seoktae Koh}
\email{skoh@itp.ac.cn} \affiliation{ Institute of Theoretical
Physics, Chinese Academy of Sciences, P.O.Box 2735, Beijing 100190,
China}

\vspace*{0.2cm}
\begin{abstract}
\baselineskip=0.6 cm
We study the effect of the Gauss-Bonnet term on vacuum decay process
in the Coleman-De Luccia formalism. The Gauss-Bonnet term has an
exponential coupling with the real scalar field, which appears in
the low energy effective action of string theories.  We calculate
numerically the instanton solution, which describes the process of
vacuum decay, and obtain the critical size of bubble. We find that
the Gauss-Bonnet term has a nontrivial effect on the false vacuum
decay, depending on the Gauss-Bonnet coefficient.
\end{abstract}

\maketitle

\section{Introduction}

Vacuum decay is an old subject in field theory~\cite{old}. Coleman
and Callan\cite{col1,col2} have shown that a quantum tunneling
process from a false vacuum to a true vacuum can be realized via the
nucleation of a true vacuum bubble in the surrounding of the false
vacuum. Furthermore, Coleman and De Luccia~\cite{col3} have found
that gravity has a significant effect on the vacuum decay process.

 In semiclassical
approximation, the decay rate per unit time per unit volume is given
by $\Gamma /V=Ae^{-B/\hbar}[1+\mathcal{O}(\hbar)]$, where the factor
$A$ has been discussed in Ref.\cite{col3,cal1} and the exponent $B$
is the difference of Euclidean actions between the instanton
solution $\phi_b$ (bounce solution) and false vacuum solution
$\phi_{F}$. Recently, some authors have discussed the vacuum decay
in the different situations such as nonminimal coupling between the
scalar field and curvature scalar~\cite{lee}, DBI action~\cite{dbi},
 and non-thin-wall limit~\cite{nonthin}, etc. Especially, a new kind
 of bounce solutions in de Sitter spacetime, which is called by
 oscillating bounce, has been found by~\cite{osci,banks}.

The finite temperature effect on the false vacuum decay process has
also been discussed by Linde {\it et. al.}\cite{temp}, where one
should look for the $O(3)$-symmetric solution due to periodicity in
the time direction with the period of inverse temperature $T^{-1}$,
instead of the $O(4)$-symmetric solution at zero temperature. The
cosmological applications of false vacuum decay process to various
inflation cosmological models have been extensively discussed in
\cite{guth}.

Recently, the so-called stringy landscape scenario~\cite{lands}
predicts that there is a big number of vacua in the effective theory
of string theories. On the other hand, a lot of astronomical
observations indicate a tiny positive cosmological constant exists
in our universe. These motivate many new investigations on the
vacuum decay. It is well-known that the higher derivative terms of
gravity naturally appear in the low energy effective action of
string theories. By  field redefinition, $R^2$ terms corrections can
be recast to a Gauss-Bonnet form~\cite{Gross,Zwie}. In particular,
the low energy theory with the Gauss-Bonnet is free of ghosts,
evading any problem with the unitarity~\cite{Zwie}.  The possible
role of such a Gauss-Bonnet term in the inflation and dark energy
models has been investigated
recently~\cite{sasa,ccd,guo,soda1,soda2,soda3,mota1,mota2}. The
instability of vacua for the Gauss-Bonnet branch in the Gauss-Bonnet
gravity is also investigated by a very recent paper~\cite{xi}.
Therefore, it is of great interest to see whether the Gauss-Bonnet
term has any effect on the vacuum decay. This is just the aim of the
present paper. We find that the Gauss-Bonnet term indeed has a
significant effect on the vacuum decay process.

Before proceeding, let us first stress the issue of stability of
vacua. In the absence of gravity, any local minimum of potential
for a scalar field can be viewed as a vacuum of the scalar field.
The vacuum with the lowest energy density called true one, while
others false vacua. Energy density of the field at the vacuum can
be positive, zero or negative, the vacuum is always classically
stable even for the case with a negative energy density if the
potential has a lower bound. When gravity appears, three cases
correspond to de Sitter, Minkowski and anti-de Sitter spacetimes,
respectively. Due to the absence of ghost in the Gauss-Bonnet
gravity, these vacua are classically stable. However, if the true
vacuum has a negative energy density, the spacetime inside bubble
is anti-de Sitter universe by quantum tunneling from a false
vacuum. The anti-de Sitter universe is unstable and will collapse.
On the other hand, Minkowski and de Sitter universe are stable.
Therefore in this paper we will not consider the case with a
negative energy density in true vacuum.

This paper is organized as follows. In Sec.~II we present the
Euclidean action, equation of motion (EoM) of the scalar field
$\phi$, as well as the Einstein equations. In Sec.~III we
numerically calculate the nucleation of a Minkowski true vacuum
from a de Sitter false vacuum. And we consider three different
Gauss-Bonnet coefficient $\alpha$ values to investigate the
Gauss-Bonnet term effect. In Sec.~IV we compute the exponent $B$
and the critical size of bubble radius analytically in thin-wall
approximation. We also mention the classical growth of the bubble.
Sec.~V includes our conclusion and discussion.

\section{Action and Equations of Motion}

We consider the following low energy effective action with an
exponential coupling between the Gauss-Bonnet term and the scalar
field
\begin{equation}
\label{1eq1} S=\int
d^{3}xdt~\sqrt{-g}\left\{\frac{1}{2\kappa^{2}}R-\frac{1}{2}\partial_{\mu}\phi\partial^{\mu}\phi-U(\phi)+\alpha
e^{\beta\phi}R^{2}_{GB}\right\},
\end{equation}
where $\kappa^{2}=8\pi G=8\pi/M^2_{pl}=8\pi \l^2_{pl}$, the
signature of the metric is $(-,+,+,+)$,
$R^{2}_{GB}=R^{2}-4R_{\mu\nu}R^{\mu\nu}+R_{\mu\nu\rho\sigma}R^{\mu\nu\rho\sigma}$,
$U(\phi)$ is the potential of the scalar field,  $\beta$ is the
coupling constant of the scalar field to the Gauss-Bonnet term and
$\alpha$ is called Gauss-Bonnet coefficient. Here we have
neglected the boundary term $S_{b}$ associated with the scalar
curvature and the Gauss-Bonnet term, because it will be canceled
in our computation.

Changing to the Euclidean signature by virtue of $t=i\eta$, we
obtain the Euclidean action
\begin{equation}
\label{1eq2} S_{E}=-\int
d^{3}xd\eta~\sqrt{g}\left\{\frac{1}{2\kappa^{2}}R-\frac{1}{2}\partial_{\mu}\phi\partial^{\mu}\phi-U(\phi)+\alpha
e^{\beta\phi}R^{2}_{GB}\right\}.
\end{equation}
Following \cite{col3}, we consider the metric of $SO(4)$-symmetry
\begin{eqnarray}
\label{1eq3}
ds_{E}^{2}&=&d\eta^{2}+\rho^{2}(\eta)d\Omega_{(3)}^{2}\nonumber\\
&=&d\eta^{2}+\rho^{2}(\eta)(d\theta^{2}+\sin^{2}\theta
d\chi^{2}+\sin^{2}\theta \sin^{2}\chi d\psi^{2}),
\end{eqnarray}
which has the curvature scalar and Gauss-Bonnet term
$~R=-\frac{6(-1+\dot{\rho}^{2}+\rho\ddot{\rho})}{\rho^{2}}$, and
$~R^{2}_{GB}=-24\frac{\ddot{\rho}}{\rho^{3}}(1-\dot{\rho}^{2})$,
respectively. Here an overdot stands for the derivative with
respect to $\eta$.  Plugging into Eq.(\ref{1eq2}), the Euclidean
action reduces to
\begin{equation}
\label{1eq4} S_{E}=2\pi^{2}\int d\eta~
\rho^{3}\left\{\frac{1}{2}\dot{\phi}^{2}+U(\phi)-\frac{3}{\kappa^{2}}\frac{(1-\dot{\rho}^{2}-\rho\ddot{\rho})}{\rho^{2}}+24\alpha
e^{\beta\phi}\frac{\ddot{\rho}}{\rho^{3}}(1-\dot{\rho}^{2})\right\},
\end{equation}
 Within the action (\ref{1eq1}), the Einstein equations
read~\cite{sasa,ccd}
\begin{eqnarray}
\label{1eq5} 0& =&
\frac{1}{\kappa^{2}}R_{\mu\nu}-\partial_{\mu}\phi\partial_{\nu}\phi+g_{\mu\nu}\left[-\frac{1}{2\kappa^{2}}R
+4\nabla^{2}f-8R^{\sigma\tau}\nabla_{\sigma}\nabla_{\tau}f+\frac{1}{2}\partial_{\rho}\phi\partial^{\rho}\phi+U(\phi)\right ]\nonumber\\
& &-4R\nabla_{\mu}\nabla_{\nu}f-8R_{\mu\nu}\nabla^{2}f
-8R_{(\mu~~~~\nu)}^{~~~\sigma\tau}\nabla_{\sigma}\nabla_{\tau}f+16R_{\sigma(\mu}\nabla^{\sigma}\nabla_{\nu)}f,
\end{eqnarray}
where $f=\alpha e^{\beta\phi}$. The $\eta\eta$ component and
$\theta\theta$ component of  Einstein equations are
\begin{eqnarray}
\label{1eq6} &&
0=\frac{1}{2}\dot{\phi}^{2}-U(\phi)+\frac{3}{\kappa^{2}}\frac{1}{\rho^{2}}(1-\dot{\rho}^{2})-24\alpha\beta
e^{\beta\phi}\dot{\phi}\frac{\dot{\rho}}{\rho^{3}}(1-\dot{\rho}^{2}),
\\
\label{1eq7} &&
0=\rho^{2}[\frac{1}{2}\dot{\phi}^{2}+U(\phi)]-\frac{1}{\kappa^{2}}(1-\dot{\rho}^{2}-2\rho\ddot{\rho})
-24\alpha\beta
e^{\beta\phi}\dot{\phi}\frac{\dot{\rho}}{\rho}(1-\dot{\rho}^{2})\nonumber\\
&&~~~~~~-24\alpha\beta
e^{\beta\phi}\dot{\phi}\dot{\rho}\ddot{\rho}+8\alpha\beta
e^{\beta\phi}(\beta\dot{\phi}^{2}+\ddot{\phi})(1-\dot{\rho}^{2}).
\end{eqnarray}
Varying the Euclidean action (\ref{1eq4}) with respect to $\phi$
yields  the EoM of $\phi$
\begin{equation}
\label{1eq8}
\ddot{\phi}+3\frac{\dot{\rho}}{\rho}\dot{\phi}-\frac{dU}{d\phi}-24\alpha\beta
e^{\beta\phi}\frac{\ddot{\rho}}{\rho^{3}}(1-\dot{\rho}^{2})=0.
\end{equation}

\section{Numerical Calculation}

\begin{figure}
\includegraphics[width=9cm]{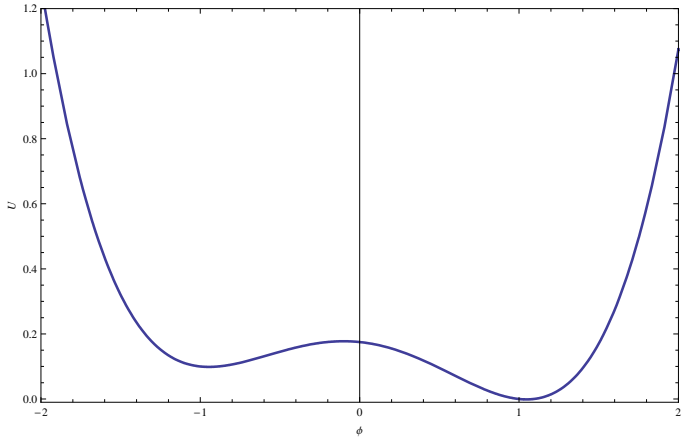}
\caption{\label{potential1}Potential profile with
$\tilde\epsilon\equiv\frac{\lambda}{\mu^{4}}\epsilon=-0.1$.}
\end{figure}

As usual, we consider the one-loop effective potential as
follows~\cite{col3}.
\begin{equation}
\label{2eq1} U(\phi)=U_{0}+{\rm 1~loop~
term}=\frac{\lambda}{8}(\phi^{2}-\mu^{2}/\lambda)^{2}+\frac{\epsilon}
{2\sqrt{\mu^{2}/\lambda}}(\phi-\sqrt{\mu^{2}/\lambda}),
\end{equation}
where $\mu$ and $\lambda$ are Higgs mass and coupling constant,
respectively, and the second term corresponds to one-loop
correction. When the correction is absent, the potential $U_0$ has
two degenerate  vacua at $\phi_{\pm}= \pm \sqrt{\mu^2/\lambda}$,
with vanishing potential. The correction term eliminates the
degeneracy. The constant $\epsilon$ stands for the potential
energy difference between two vacua.  The correction does not
change the value of the potential at $\phi_+$, but a positive
$\epsilon$ shifts the potential at $\phi_-$ down and a negative
$\epsilon$ shifts the potential up.  In Fig.~\ref{potential1} we
plot the potential profile with a negative $\epsilon$.  In that
case, the vacuum at $\phi_T=\sqrt{\mu^2/\lambda}$ corresponds to a
Minkowski one with vanishing potential, while the vacuum at
$\phi_F=-\sqrt{\mu^2/\lambda}$ is a false vacuum, corresponding to
a de Sitter vacuum in gravity theory. If we take $\epsilon$ to be
a positive one, then the vacuum at $\phi_+$ turns to be a false
vacuum with vanishing potential, while the vacuum at $\phi_-$ is a
true one with a negative potential, corresponding to an anti-de
Sitter vacuum, once gravity is taken into account.

We see from (\ref{1eq8}) that when the GB term is absent, the vacuum
structure of the potential is completely determined by the potential
itself. However, when the GB term appears, the vacuum becomes to be
determined by the third and fourth terms in (\ref{1eq8}). We can
define an effective potential
\begin{equation}
\label{2eq2}
 U_{eff}= U + 24 \alpha e^{\beta \phi}\ddot{\rho} (1-\dot
\rho^2)/\rho^3.
\end{equation}
Then new vacuum is determined by $\left. dU_{eff}/d\phi\right
|_{\phi^{new}_v}=0$. To be more clear, for a Minkowski vacuum, one
has $~\rho=\eta$,$~U_{eff}=U$, i.e., GB term takes no effect on
Minkowski vacuum; for a de Sitter vacuum,
$~\rho=\Lambda_{1}\sin\frac{\eta}{\Lambda_{1}}$, ({\ref{1eq6}})
and ({\ref{2eq2}}) give
$\Lambda_{1}=(\frac{3}{\kappa^{2}U(\phi^{new}_v)})^{1/2}$ and
$~U_{eff}=U-24\frac{\alpha}{\Lambda_{1}^{4}}e^{\beta\phi^{new}_v}$,
respectively, where $\phi^{new}_v$ represents new false vacuum
value $\phi^{new}_{F}$ or new true vacuum value $\phi^{new}_{T}$
which is determined by $\left. dU_{eff}/d\phi\right
|_{\phi^{new}_v}=0$; similarly, for an anti-de Sitter vacuum,
$~\rho=\Lambda_{2}\sinh\frac{\eta}{\Lambda_{2}}$,
$\Lambda_{2}=(-\frac{3}{\kappa^{2}U(\phi^{new}_v)})^{1/2}$ and
$~U_{eff}=U-24\frac{\alpha}{\Lambda_{2}^{4}}e^{\beta\phi^{new}_v}$,
in which $\phi^{new}_v$ is also given by $\left.
dU_{eff}/d\phi\right |_{\phi^{new}_v}=0$. As a summary, in
Minkowski case, GB term keeps potential $U(\phi)$ unchanged; in
both de Sitter and anti-de Sitter case, GB term shifts the vacuum
value of scalar field from an old one to a new one ($\phi^{old}_v
\rightarrow\phi^{new}_v$), meanwhile, the original potential
energy is replaced by an effective potential ($U\rightarrow
U_{eff}$), in which $\phi^{old}_v$ ($\phi^{new}_v$) is computed by
$\left. dU/d\phi\right |_{\phi^{old}_v}=0$ ($\left.
dU_{eff}/d\phi\right |_{\phi^{new}_v}=0$).

\begin{figure}
\includegraphics[width=9cm]{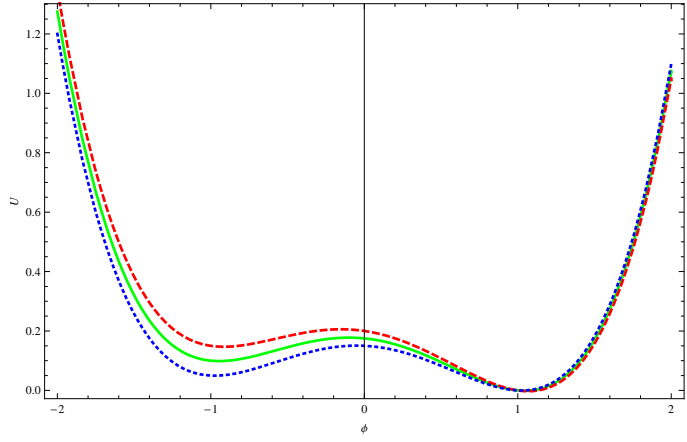}
\caption{\label{potetntial3}Potential profile with different
$\tilde\epsilon$.  The solid line corresponds to
$\tilde\epsilon=-0.1$,
 the dashed line to
$\tilde\epsilon= -0.15$ and the dotted line to $\tilde\epsilon
=-0.05$.}
\end{figure}

We conclude, from above analysis, that, at a de Sitter vacuum,
$~U_{eff}=U-24\frac{\alpha}{\Lambda_{1}^{4}}e^{\beta\phi^{new}_v}$,
a positive $\alpha$ is analogous to the case to decrease
$|\epsilon|$ in the Einstein gravity, while a negative $\alpha$
leads to an opposite effect (See Fig.~\ref{potetntial3}); at anti-de
Sitter vacuum, the cases are reverse,
$~U_{eff}=-|U|-24\frac{\alpha}{\Lambda_{2}^{4}}e^{\beta\phi^{new}_v}$.
That is, a positive $\alpha$ corresponds to increasing $|\epsilon|$,
while a negative $\alpha$ to decreasing $|\epsilon|$; at Minkowski
vacuum, the GB term takes no effect.

In addition, we must point out that the shifts of vacuum potential
energy due to the appearance of GB term can not change the topology
of the original spacetime manifold. For example, if the original
potential energy is definitely positive ($U(\phi)>0$), we can only
get new de Sitter solution. Neither anti-de Sitter solution nor
Minkowski solution are permitted in the new effective potential.
That is to say, a de Sitter vacuum could not become a Minkowski
vacuum or an anti-de Sitter vacuum due to the Gauss-Bonnet term.
This can be seen from (\ref{1eq6}).  The same holds for the
Minkowski vacua case and anti-de Sitter vacua case.  Note that in
the non-minimal coupling case~\cite{lee}, such a change is possible.
That is, in that case, a true vacuum could turn to be a false vacuum
due to the non-minimal coupling.

In the low energy effective action of string theories, the GB term
can be parameterized as~\cite{guo}
\begin{equation}
\label{2eq3} \alpha
e^{\beta\phi}R^{2}_{GB}=-\frac{1}{2}\frac{\alpha^{\prime}}{\kappa^{2}}\gamma
e^{-\phi/M_{pl}}R^{2}_{GB},
\end{equation}
where $M_{pl}$ is the Planck mass, and $\alpha'=\l^2_{s}$ is the
string slop. That is to say, we take
$\alpha=-\frac{1}{16\pi}\gamma(\frac{\l_{s}}{\l_{pl}})^2$, and
$\beta =-1/M_{pl}$.  Note that $\gamma=-\frac{1}{4},-\frac{1}{8}$,
and $0$, correspond to the cases in the low energy effective
theory of bosonic, heterotic, and type II superstring theory,
respectively. In order not to make confusion, we should stress
here that in fact, there is no quadratic correction in type II
superstring theory.

For numerical calculations, we make the following rescalings so that
these quantities become dimensionless
$$ ~\tilde{\phi}=\sqrt{\frac{\lambda}{\mu^{2}}}\phi,\ \ \tilde{\eta}=\mu\eta,
~\tilde{\epsilon}=\frac{\lambda}{\mu^{4}}\epsilon,\ \
\tilde{\rho}=\mu\rho,\ \
~\tilde{\beta}=\sqrt{\frac{\mu^{2}}{\lambda}}\beta,\ \
~\tilde{\alpha}=\lambda\alpha,
~\tilde{U}(\tilde{\phi})=\frac{\lambda}{\mu^{4}}U(\phi),~\tilde{\kappa}^{2}=\frac{\mu^{2}}{\lambda}\kappa^{2}.$$
In principle we can discuss the quantum tunneling among various
vacua as was done in Ref.\cite{lee}. Note that if the true vacuum
is an anti-de Sitter one, the resulting spacetime is dynamically
unstable as stressed in Introduction. To demonstrate the role of
the GB term in the vacuum decay, here we focus on the case in
which a de Sitter vacuum decays into a Minkowski vacuum only by
using the potential (\ref{2eq1}). In the following numerical
calculation we fix $\tilde{\epsilon}=-0.1$ as well as
$\tilde{\kappa}^{2}=0.1$ and keep $\tilde{\alpha}$ free, which
mainly depends on the string length scale. In convenience, we will
drop the tilde symbol in the following when we use dimensionless
quantities.

\begin{figure}
\includegraphics[width=9cm]{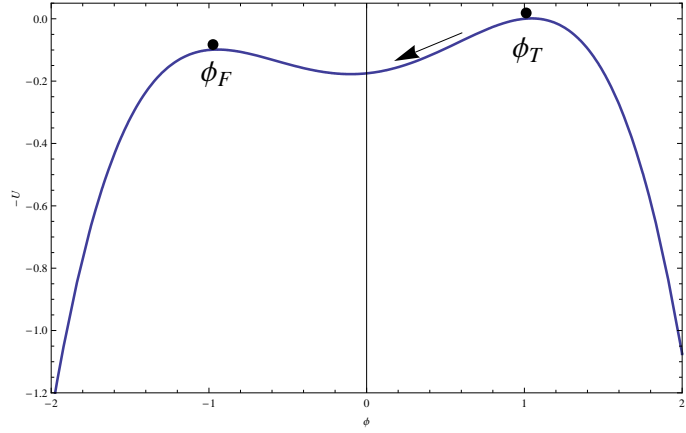}
\caption{\label{potetntial2}Reversed potential profile with
$\tilde\epsilon=-0.1$.}
\end{figure}

\begin{figure}
\includegraphics[width=9cm]{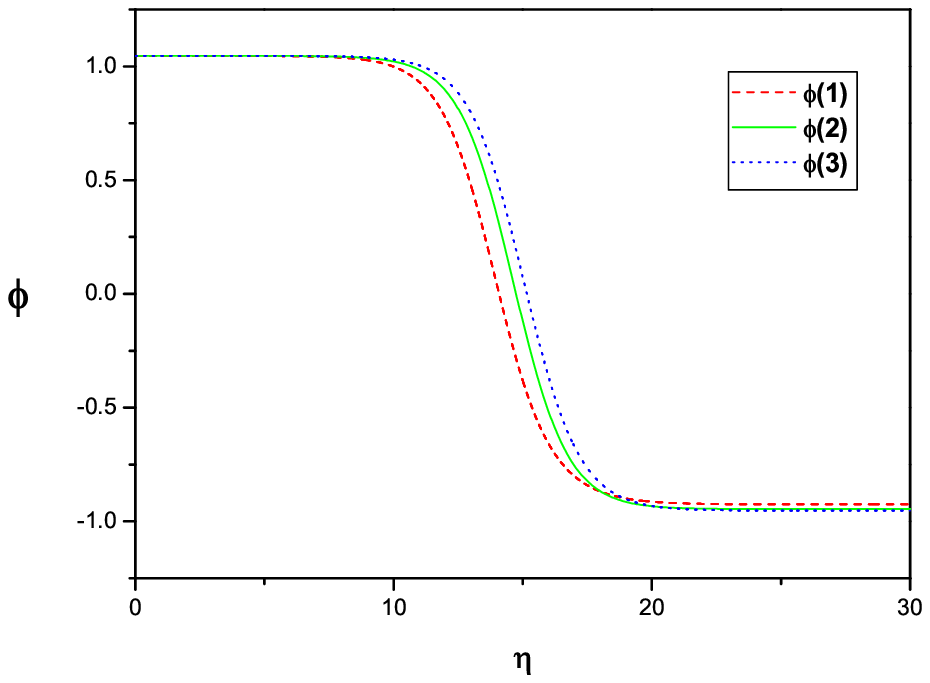}
\caption{\label{Graph1} $\phi$ profile v.s. $\eta$. The solid,
dashed and dotted curves correspond to $\alpha=0$, $-10^{3}$, and
$300$, respectively.}
\end{figure}
\begin{figure}
\includegraphics[width=9cm]{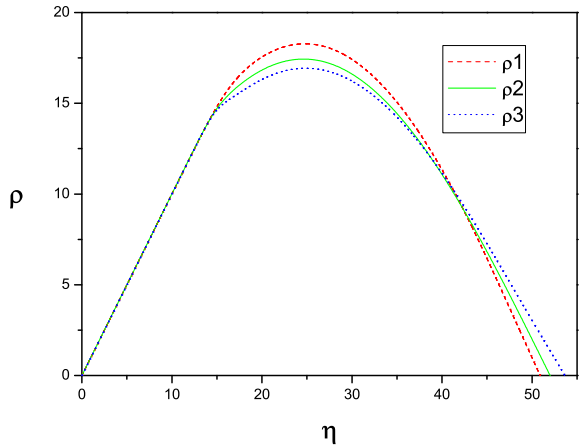}
\caption{\label{Graph2} $\rho$ profile v.s. $\eta$. The solid,
dashed and dotted curves correspond to $\alpha=0$, $-10^{3}$, and
$300$, respectively.}
\end{figure}

As Coleman~\cite{col1} has demonstrated that, at semiclassical
level, the quantum tunneling from a false vacuum to a true vacuum in
Lorentzian spacetime is analogous to the problem that a classical
particle rolls down from a higher peak to a lower peak in the minus
potential in Euclidean spacetime. That is to say, if the particle is
released at rest at a proper position near $\phi_{T}$ it will come
to rest at $\eta=\eta_{max}$ at $\phi_{F}$. (In de Sitter spacetime
there exists a maximum $\eta_{max}$, so we let the particle come to
rest at $\eta_{max}$, while in Minkowski or anti-de Sitter spacetime
there does not exist such $\eta_{max}$, so in those cases, we will
let particle reach the lower peak when $\eta$ goes to infinity.)
Thus, we can solve the EoM of $\phi$ and Einstein equations as a
boundary value problem
\begin{equation}
\label{2eq4}
 ~\left. \frac{d\phi}{d\eta}\right |_{\eta=0}=0,
~\left. \phi\right |_{\eta_{max}}=\phi_{F},
~\dot{\rho}|_{\eta=0}=\dot{\rho}_{0},~\rho|_{\eta=0}=0.
\end{equation}

Among Eqs.(\ref{1eq6}-\ref{1eq8}) only two equations are
independent, because the three equations are related by Bianchi
identity. We combine the EoM of $\phi$ and $\theta\theta$ component
equation to solve the problem and choose the $\eta\eta$ component
equation as a constraint which, on one hand, constrains our
solution, on the other hand, determines the initial value of
$\dot{\rho}$, i.e., the value of $\dot{\rho}_{0}$.

Although in string theories $\alpha$ is always positive definite, we
still consider a negative $\alpha$ case because that case
corresponds to shift up the false vacuum (de Sitter vacuum). In this
paper, we take three different $\alpha$ values to see the effect of
GB
 term. Case 1: $\alpha=0$, which corresponds to the case without the GB term;
case 2: $\alpha=-10^{3}$, which corresponds to de Sitter vacuum
shifting up; case 3: $\alpha=300$,  corresponding to de Sitter
vacuum shifting down but still a de Sitter vacuum. The numerical
results are shown in Fig.~\ref{Graph1} and Fig.~\ref{Graph2}. We can
see from Fig.~\ref{Graph1} that when $\alpha$ is negative, quantum
tunneling can happen at smaller $\eta$, compared to the case with
 $\alpha =0$, while it can occur at larger $\eta$ for a
positive $\alpha$.  We can also find that GB term makes the false
vacua values of scalar field $\phi^{new}_F$ (dashed and dotted
curves) have tiny shifts compared with $\phi^{old}_F$ (solid curve),
however, GB term leaves $\phi_T$ (Minkowski vacuum) unchanged (three
curves coincide with each other at true vacuum). As a qualitative
approximation, if neglect the thickness of the wall of the bubble,
we can write down the metric of the whole spacetime
\begin{equation}
\label{2eq5} \left\{\begin{array} {l@{\quad~\quad}l}
\rho=\eta,&(0\leq\eta\leq\eta_w),\\\rho=\Lambda_{eff}
\sin(\eta/\Lambda_{eff}),&(\eta_w\leq\eta\leq\eta_{max}),\end{array}\right.
\end{equation}
where $\eta_w$ represents for the position of the wall, and
$\eta_{max}=\pi \Lambda_{eff}$. Crossing the the wall at $\eta_w$,
spapcetime metric and matter field should be matched by the
conjunction conditions~\cite{jc1,jc2}.  Of course, this is just a
rough approximation. In fact, due to the existence of the wall
supported by the scalar field, the spacetime metric and matter field
should be smoothly continuous across the wall. In that case, the
conjunction condition is not necessary.

 We can
see from figures that as we analyzed above, indeed, in the case we
considered, the effect of the GB term is qualitatively equivalent to
changing the potential difference $\epsilon$ between two vacua in
Einstein gravity. A negative $\alpha$ term corresponds to increasing
the potential difference, and a positive $\alpha$ term results in an
opposite effect.

\section{Thin Wall Approximation}

In general it is impossible to find an analytic solution to
describe the process of vacuum decay, even in the case of absence
of gravity. But as shown in \cite{col3}, it is possible to find an
approximate solution in the thin wall approximation. The so-called
thin wall approximation means that the thickness of the wall is
quite small compared to the size of the bubble. This happens
$|\tilde \epsilon| \ll 1$ for the potential (\ref{2eq1}), which
implies that the energy difference between the false vacuum and
true vacuum is small compared to the height of the barrier between
these two vacua. As shown in \cite{col3}, in the thin wall
approximation, the thickness of the wall is of ${\cal
O}(\mu^{-1})$, while the size of the bubble is proportional to the
inverse of the energy difference of two vacua.

In this section we will calculate the critical size of the bubble
in the thin wall approximation. The critical size of the bubble is
determined by requiring that the Euclidean action difference be
stationary, between the bounce (instanton) solution and the false
vacuum solution.  If the radius of the bubble after nucleation is
smaller than the critical radius, the bubble can not grow up,
because the decrement of volume energy is less than the increment
of surface energy. That is to say, only when
$\bar{\rho}\geq\bar{\rho}_{c}$ the nucleated bubble can grow up.
Here $\bar \rho$ denotes the size of the bubble.  Now we calculate
the critical radius $\bar\rho_c$ of the bubble. Note that $B$ is
the action difference between the bounce solution $\phi_b$ and the
constant false vacuum solution $\phi_F$. Then the critical radius
of the bubble is determined by $dB/d\bar \rho
|_{\bar\rho=\bar\rho_c}=0$.

Following~\cite{col3}, we calculate the Euclidean action $S_{E}$
by dividing the solution into three parts: inside the wall
$S_{E}^{i}$, the wall  $S_{E}^{w}$, and outside the wall
$S_{E}^{o}$. Outside the wall,  $S_{E}^{o}(\phi_{b})$ and
$S_{E}^{o}(\phi_{F})$ cancels each other, so we have $B$ as
\begin{equation}
\label{3eq1} B=S_{E}^{i+w}(\phi_{b})-S_{E}^{i+w}(\phi_{F}).
\end{equation}
The Euclidean action reads
\begin{equation}
\label{3eq2} S_{E}=2\pi^{2}\int d\eta~
\rho^{3}\left\{\frac{1}{2}\dot{\phi}^{2}+U(\phi)-\frac{3}{\kappa^{2}}
\frac{(1-\dot{\rho}^{2})}{\rho^{2}}+\frac{3\ddot{\rho}}{\kappa^{2}\rho}+24\alpha
e^{\beta\phi}\frac{\ddot{\rho}}{\rho^{3}}(1-\dot{\rho}^{2})\right\}.
\end{equation}

On the wall, in the thin wall approximation, the second and fourth
terms in the equation of motion for $\phi$ in (\ref{1eq8}) can be
neglected (we will discuss this below) . This implies that the
last term in (\ref{3eq2}) can be neglected as well on the wall. By
virtue of integration by parts and Einstein equation (\ref{1eq6}),
the Euclidean action is changed to
\begin{equation} \label{3eq23}
S_{E}(\phi)=4\pi^{2}\int_{\eta_1}^{\eta_{3}}d\eta~\left(\rho^3U-\frac{3}{\kappa^2}\rho\right),
\end{equation}
where we drop the surface terms because those terms are always
cancelled when we calculate the difference of the Euclidean action
($B$). The action (\ref{3eq23}) can be further approximated as
\begin{equation} \label{3eq25}
S^w_{E}(\phi_b)=4\pi^{2}\int_{\eta_1}^{\eta_{3}}d\eta~\left(\bar\rho^3U_0(\phi_b)-\frac{3}
{\kappa^2}\bar\rho\right),
\end{equation}
where we have used $\rho\simeq\bar\rho$, and $U(\phi)\simeq
U_0(\phi)$ (thin wall approximation)~\cite{col3}.

For the false vacuum solution, on the wall, we can also replace
$\rho$ by $\bar\rho$, and $U(\phi)$ by $U_0(\phi)$, then the
Euclidean action reads
\begin{equation}
\label{3eq26}
S_{E}^{w}(\phi_{F})=4\pi^{2}\int_{\eta_{1}}^{\eta_{3}}d\eta~
\left(\bar{\rho}^{3}U_{0}(\phi_F)-\frac{3}{\kappa^{2}}\bar{\rho}^{2}\right).
\end{equation}
\begin{figure}
\includegraphics[width=9cm]{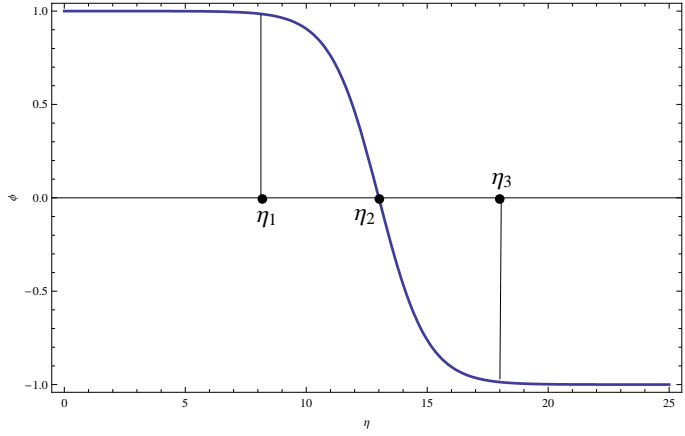}
\caption{\label{instanton} The instanton solution profile $\phi$
v.s. $\eta$.}
\end{figure}
Thus we have
\begin{eqnarray}
\label{3eq6}
B^{w}& \equiv &S_{E}^{w}(\phi_{b})-S_{E}^{w}(\phi_{F})\nonumber\\
     &\approx &2\pi^{2}\bar{\rho}^{3}S_{1}
\end{eqnarray}
where
$S_{1}\equiv\int_{\eta_{1}}^{\eta_{3}}d\eta~2[U_{0}(\phi)-U_{0}(\phi_{F})]$.
This result indicates that the Gauss-Bonnet term has no
contribution to the wall part of the Euclidean action difference.
This is an expected result since one can see from (\ref{3eq6})
that even the Hilbert-Einstein term has no contribution by noting
that the form (\ref{3eq6}) is completely the same as the case
without gravity~\cite{col3}.

Inside the wall, the Euclidean action reads
\begin{equation}
\label{3eq7}
S^i_{E}(\phi)=4\pi^{2}\int_0^{\eta_{2}}d\eta~\left(\rho^3U(\phi)-\frac{3}{\kappa^2}\rho\right)
+96\pi^2\alpha\int_0^{\eta_{2}}d\eta~e^{\beta\phi}\dot\rho^2\ddot\rho,
\end{equation}
and the Einstein equations are
\begin{eqnarray}
\label{3eq8} &&
\dot{\rho}^{2}=1-\frac{1}{3}\kappa^{2}U(\phi)\rho^{2},\\
\label{3eq9} &&
\ddot{\rho}=\frac{1}{2\rho}[1-\dot{\rho}^{2}-\kappa^{2}U(\phi)\rho^{2}]=-\frac{1}{3}\kappa^{2}U(\phi)\rho.
\end{eqnarray}
By virtue of Eqs. (\ref{3eq8}) and (\ref{3eq9}), the Euclidean
action becomes
\begin{eqnarray}
\label{3eq10} S_{E}^{i}(\phi_{b})
&=&-\frac{12\pi^2}{\kappa^2}\int_0^{\bar{\rho}}d\rho~\rho\sqrt{1-\frac{\kappa^2}{3}U(\phi_T)\rho^{2}}\nonumber\\
&=&-\frac{6\pi^{2}\bar{\rho}^{2}}{\kappa^{2}},
\end{eqnarray}
where we have used the fact $U(\phi_{T})=0$,
$\phi_{b}\simeq\phi_{T}$. It is so because inside the wall, the
true vacuum is a Minkowski spacetime, where the Gauss-Bonnet term
has no contribution. On the other hand, the Euclidean action from
the false vacuum is
\begin{eqnarray}
\label{3eq11} S_{E}^{i}(\phi_{F})
&=&-\frac{12\pi^{2}}{\kappa^{2}}\int_0^{\bar{\rho}}d\rho~\rho\sqrt{1-\frac{1}{3}D_2\rho^{2}}
-32\pi^2\alpha\int_0^{\bar{\rho}}d\rho~e^{\beta\phi_F}D_2\rho\sqrt{1-\frac{1}{3}D_2\rho^{2}}\nonumber\\
&=&\left(\frac{12\pi^{2}}{\kappa^{2}D_2}+32\pi^2\alpha
e^{\beta\phi_F}\right)\left[(1-\frac{1}{3}D_{2}\bar{\rho}^{2})^{3/2}-1\right].
\end{eqnarray}
We have defined the position of the wall as
$\bar{\rho}=\rho(\eta_{2}) \approx \rho(\eta_{1})$ because of the
thin wall approximation, and $D_{2}\equiv\kappa^{2}U(\phi_{F})$.
As a result, we get the contribution $B^{i}$ inside the wall
{\begin{eqnarray} \label{3eq12}
B^{i}& \equiv &S_{E}^{i}(\phi_{b})-S_{E}^{i}(\phi_{F})\nonumber\\
&=&-\frac{6\pi^{2}\bar{\rho}^{2}}{\kappa^{2}}-\left(\frac{12\pi^{2}}{\kappa^{2}D_2}+32\pi^2\alpha
e^{\beta\phi_F}\right)\left[(1-\frac{1}{3}D_{2}\bar{\rho}^{2})^{3/2}-1\right].
\end{eqnarray}
\begin{figure}
\includegraphics[width=9cm]{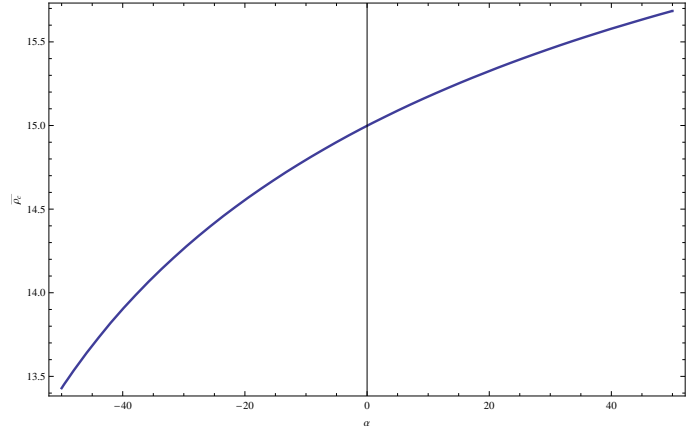}
\caption{\label{alpha-rhoc1} The critical size of the bubble
$\bar{\rho}_{c}$ v.s. the GB coefficient $\alpha$.}
\end{figure}
Finally we obtain the total Euclidean action difference $B$ as
\begin{eqnarray} \label{3eq13} B
=2\pi^{2}\bar{\rho}^{3}S_{1}-\frac{6\pi^{2}\bar{\rho}^{2}}{\kappa^{2}}-\left(\frac{12\pi^{2}}{\kappa^{2}D_2}+32\pi^2\alpha
e^{\beta\phi_F}\right)\left[(1-\frac{1}{3}D_{2}\bar{\rho}^{2})^{3/2}-1\right].
\end{eqnarray}
The critical size of bubble is determined  through
\begin{equation}
\label{3eq14} \left. \frac{dB}{d\bar{\rho}}\right |_
{\bar{\rho}=\bar{\rho}_{c}}=0.
\end{equation}
Substituting Eq. (\ref{3eq13}) into the above equation yields
\begin{equation} \label{3eq15}
3S_1\bar\rho_c-\frac{6}{\kappa^2}+D_3(1-\frac{1}{3}D_2\bar\rho_c^2)^{1/2}=0,
\end{equation}
where $D_3=6/\kappa^2+16\alpha D_2 e^{\beta\phi_F}.$ As
$\alpha=0$, Eq.(\ref{3eq15}) leads to the Coleman-de Luccia's
result~\cite{col3}:
\begin{equation}
\label{3eq16}
\bar{\rho}_{c}=\frac{12S_{1}}{4U(\phi_{F})+3\kappa^{2}S_{1}^{2}}.
\end{equation}
If $\alpha\neq0$, one has
\begin{equation}
\label{3eq17}
\bar{\rho}_{c}=\frac{36S_1+\sqrt{(36S_1)^2-4(36-\kappa^4D_3^2)
(\frac{1}{3}D_2D_3^2+9S_1^2)}}{2\kappa^2(\frac{1}{3}D_2D_3^2+9S_1^2)}.
\end{equation}
In Fig.~\ref{alpha-rhoc1} we plot the critical size of the bubble
versus the GB coefficient $\alpha$.  As expected, when $\alpha<0$,
the critical size of bubble becomes smaller, which implies that
bubble nucleation becomes easier, and vice versa.  These
analytical results are consistent with our previous numerical
calculation.

Before we turn to the issue on the growth of the bubble, we
discuss the validity of the thin-wall approximation. In order to
get an instant solution over the wall, the second and the fourth
terms in Eq. (\ref{1eq8}) have been neglected,  i.e., one has
$\dot{\bar\rho}/\bar\rho \ll 1$ and
$(1-\dot{\bar\rho}^2)/\bar\rho^2 \ll 1$. Now we justify this. By
 Eq. (\ref{1eq6}),
\begin{equation} \label{3eq27}
\frac{\dot{\bar\rho}^2}{\bar\rho^2}+8\alpha\beta\kappa^2\dot\phi
e^{\beta\phi}\frac{\dot{\bar\rho}}{\bar\rho}\frac{(1-\dot{\bar\rho}^2)}{\bar\rho^2}=
\frac{1}{\bar\rho^2}+\frac{\kappa^2}{3}\left(\dot\phi^2/2-U\right).
\end{equation}
The left hand side of this equation is certainly small if both terms
on the right are small. $1/\bar\rho^2 \ll 1$ is required as in the
absence of gravity~\cite{col1}. In fact, this is also a natural
consequence of large bubble.  The quantity in parentheses on the
righthand side can be viewed as total energy of the particle in
inverse potential. The total energy vanishes inside the wall and has
absolute value $|\epsilon|$ outside the wall. So, the absolute value
of the energy density over the wall must be smaller than
$|\epsilon|$, i.e. the absolute value of the second term on the
righthand side is smaller than
$1/\Lambda^2\equiv\kappa^2|\epsilon|/3$. Besides that, the
continuity condition of the metric on the wall gives
$\bar\rho/\Lambda \ll 1$. In conclusion, the thin-wall approximation
is valid when $\Lambda \gg \bar\rho \gg  \mu^{-1}$.

The classical growth of the bubble after its quantum nucleation is
similar to the result in Ref.\cite{lee}. For completeness, here we
just briefly mention main results.  We obtain the Lorentzian
solution from the Euclidean solution by employing the analytic
continuation in (\ref{1eq3})
\begin{equation}
\label{3eq18} \theta\rightarrow i\theta+\frac{\pi}{2}.
\end{equation}
And then transform the coordinate into the static spherically
symmetric coordinate by applying following coordinate transformation
\begin{eqnarray}
\label{3eq19} & &r=\eta\cosh\theta,~t=\eta\sinh\theta,\\ &
&r=\Lambda_{1}\sin\frac{\eta}{\Lambda_{1}}\cosh\theta,~t=\frac{\Lambda_{1}}{2}
\ln\frac{\cos\frac{\eta}{\Lambda_{1}}+\sin\frac{\eta}{\Lambda_{1}}\sinh\theta}
{\cos\frac{\eta}{\Lambda_{1}}-\sin\frac{\eta}{\Lambda_{1}}\sinh\theta}.
\end{eqnarray}
Here the first line corresponds to the Minkowski true vacuum, while
the second line corresponds to the de Sitter false vacuum. After
these transformations, we obtain a Minkowski spacetime inside the
bubble
\begin{equation}
\label{3eq20} ds^{2}=-dt^{2}+dr^{2}+r^{2}( d\chi^{2}+\sin^{2}\chi
d\psi^{2}),
\end{equation}
 while outside the bubble, we have a de Sitter space
 \begin{equation}
ds^{2}=-(1-\frac{r^{2}}{\Lambda_{1}^{2}})dt^{2}+\frac{dr^{2}}{1-\frac{r^{2}}{\Lambda_{1}^{2}}}
+r^{2}(d\chi^{2}+\sin^{2}\chi d\psi^{2}).
\end{equation}
The proper velocity of the bubble wall observed by an observer
outside the wall (spacelike observer) is
\begin{equation}
\label{3eq21} \frac{dr}{d\tau}=\sqrt{\frac{r^{2}}{\eta_{c}^{2}}-1},
\end{equation}
where the measure of proper time equals to
$d\tau=\sqrt{dt^{2}-dr^{2}}=\eta_{c}d\theta$, and $\eta_{c}$ is a
constant value very closed to $\bar{\eta}$.

\section{Conclusions}

In this paper we have investigated the effect of a Gauss-Bonnet term
on vacuum decay process of a scalar field. The Gauss-Bonnet term has
an exponential coupling with the scalar field. Such a coupling
 appears in the low energy effective action of some string
theories. We found that the Gauss-Bonnet term could change the
vacuum structure of the scalar field but could not change the
topology of the original spacetime manifold, i.e., a de Sitter
vacuum could become a new de Sitter vacuum but could not become a
Minkowski vacuum or an anti-de Sitter vacuum if the potential $U$ is
positive definite. Similar case happens in an original anti-de
Sitter vacuum. As to an original Minkowski vacuum, Gauss-Bonnet term
takes no effect, so the Minkowski vacuum remains unchanged.

Concretely, in this paper, we considered the effect of the
Gauss-Bonnet term on a de Sitter vacuum decaying into a Minkowski
vacuum. In this case, the Gauss-Bonnet term shifts the de Sitter
vacuum up or down depending on a negative or positive Gauss-Bonnet
coefficient $\alpha$, and keeps the Minkowski vacuum unchanged. We
calculated numerically the instanton solution with different
Gauss-Bonnet coefficient, and found that the effect of the
Gauss-Bonnet term is qualitatively equivalent to increasing or
decreasing the potential energy difference between the false vacuum
and the true vacuum, which makes the bubble nucleation easier or
harder.

We also computed the exponent coefficient $B$ in the decay rate and
the critical radius of the bubble in the thin-wall approximation. If
the radius of nucleated bubble is smaller than its critical size,
the bubble will shrink,  while if it is larger than the critical
size, it can grow up after quantum nucleation. We found that a
negative Gauss-Bonnet coefficient $\alpha$ leads to a smaller
critical radius, and a positive $\alpha$ to a larger critical
radius. That is to say, a negative $\alpha$ makes bubble nucleation
easier, while a positive $\alpha$ makes bubble nucleation harder,
which is consistent with our numerical calculations. In this paper,
we investigated the vacuum decay from a de Sitter vacuum to a
Minkowski vacuum. We expect the Gauss-Bonnet term has a similar
effect to other decay processes.

\begin{acknowledgments}
BH thanks H. Dong for lots of good suggestions and L. M. Cao for
many useful discussions. SK thanks W. Lee for many useful
discussions. This work was supported in part by a grant from Chinese
Academy of Sciences, grants from NSFC with No. 10325525 and No.
90403029.
\end{acknowledgments}

\end{document}